\documentclass[12pt,preprint]{aastex}
\usepackage{natbib}

\def\hmol{H$_{\rm 2}$\/}

\def\kms{km s$^{-1}$}
\def\arcsec{$"$}

\shorttitle{HD in DLA}
\shortauthors{Oliveira et al.}

\begin{document}

\title{HST/COS Detection of Deuterated Molecular Hydrogen in a DLA at $z = 0.18$}

\author{Cristina M. Oliveira}
\affil{Space Telescope Science Institute, Baltimore, MD 21218}
\email{oliveira@stsci.edu}

\author{Kenneth R. Sembach}
\affil{Space Telescope Science Institute, Baltimore, MD 21218}

\author{Jason Tumlinson}
\affil{Space Telescope Science Institute, Baltimore, MD 21218}

\author{John O'Meara}
\affil{Saint Michaels College, Colchester, VT 05439}

\and

\author{Christopher Thom}
\affil{Space Telescope Science Institute, Baltimore, MD 21218}

\begin{abstract}

We report on the detection of deuterated molecular hydrogen, HD, at $z = 0.18$. HD and \hmol~are detected in HST/COS data of a low metallicity ($Z \sim 0.07Z_\odot$) damped Ly$\alpha$ system at $z = 0.18562$ toward QSO B\,0120$-$28,  with log $N$(H\,I) = 20.50 $\pm$ 0.10.  Four absorption components are clearly resolved in H$_{\rm 2}$ while two components are resolved in HD; the bulk of the molecular hydrogen is associated with the components traced by HD. We find total column densities log $N$(HD) = 14.82 $\pm$ 0.15 and log $N$(\hmol) = 20.00 $\pm$ 0.10. This system has a high molecular fraction, $f$(\hmol) = 0.39 $\pm$ 0.10 and a low HD to \hmol~ratio, log (HD/2\hmol) $= -5.5 \pm 0.2$ dex. The excitation temperature, $T_{01} = 65 \pm 2$ K, in the component containing the bulk of the molecular gas is lower than in other DLAs. These properties are unlike those in other higher redshift DLA systems  known to contain HD, but are consistent with what is observed in dense clouds in the Milky Way. 

\end{abstract}


\section{Introduction}

DLA and sub-DLA systems offer a unique laboratory to study the evolution of the ISM in galaxies over a wide range of redshifts. Many studies have been published for DLAs at redshifts that can be observed from the ground. However, little is known about lower redshift DLAs,  particularly their molecular content. The atmospheric cutoff of light below 3000 \AA~makes ground based observing of the Ly$\alpha$ line impossible for $z < 1.5$. In addition, N(\hmol) cannot be determined from UV lines using ground-based data for $z < 1.7$, so space-based spectra are needed to determine the atomic and molecular hydrogen column densities at ultraviolet wavelengths. 

 \hmol~is only detected in 10 -- 20\% of DLA and sub-DLAs, making it difficult to study the cold ISM of these systems \citep{Battisti2012, Noterdaeme2008,Petitjean2006,Ledoux2003}.  In high redshift DLAs and sub-DLAs, \hmol~is detected in systems with high metallicity and  high gas-phase elemental depletion, and small molecular fractions \citep{Ledoux2003,Noterdaeme2008}. The molecular fractions are typically lower than at comparable N(H) in the Milky Way, but closer to values seen in the Magellanic Clouds \citep{Tumlinson2002}. With the Cosmic Origins Spectrograph (COS) on the {\it Hubble Space Telescope}, we now have access to DLAs and sub-DLAs at a redshift range $z < 0.5$, an epoch which spans $\sim$40\% of the age of the universe. Results for the first DLA and sub-DLA systems detected at $z < 0.5$ with COS have been reported by \citet{Meiring2011} and \citet{Battisti2012}.

In the Milky Way ISM, HD is detected in dense self-shielded clouds along reddened sightlines with typical values of HD/2\hmol~between a few times $10^{-7}$ and a few times $10^{-6}$ \citep{Lacour2005}.
 The formation of HD occurs primarily via the \hmol$~+$ D$^{+} =$ HD + H$^{+}$ ion-neutral reaction, and in diffuse environments HD is destroyed mainly through photodissociation by UV photons. Studying \hmol~and HD at high redshifts allows us to probe the physical properties of cool gas in systems where conditions might be very different from those in the local universe. Even though HD is a trace molecule it is sensitive to the local radiation field conditions, dust properties, and density. HD can also be a significant gas coolant at low density in low-metallicity galaxies \citep{Tumlinson2010} and understanding its chemistry and the environments in which it is formed has implications for the chemistry of dense clouds everywhere.

For the 10 -- 20\% of DLA and sub-DLAs for which \hmol~has been detected, only a handful contain absorption by deuterated molecular hydrogen, HD. 
The six detections of HD beyond the Milky Way that have been reported in the literature are all at redshifts $z > 1.7$. The HD/2\hmol~ratio ranges from log(HD/2\hmol) $= -5.03 \pm 0.10$ in a sub-DLA system at $z =$ 2.69 \citep{Noterdaeme2010} to log (HD/2\hmol) $= -4.10 \pm~^{0.22}_{0.24}$ in a sub-DLA at $z =$ 2.0594 \citep{Tumlinson2010} (see discussion in \S \ref{discussion}).

Here we report on the first detection of HD at $z < 1.7$ beyond the Milky Way galaxy. This absorption system seems to have very different characteristics from systems at higher redshift known to contain HD.

\section{The B0120$-$28 QSO and the $z_{\rm abs} =$ 0.18562 DLA}

The $z_{\rm abs} =$ 0.18562 DLA was serendipitously discovered in the HST/COS spectra of QSO B\,0129$-$28 ($z_{\rm em}$ = 0.434) from the Cycle 18 program "Probing the Ionized Gas in the Magellanic Stream" (12204, PI: Thom). QSO B\,0120$-$28 was observed on June 29, 2011 with the G130M (2 orbits) and G160M (3 orbits) gratings, providing uninterrupted wavelength coverage between 1135 and 1798 \AA. These wavelengths cover the 960 -- 1500 \AA~region in the rest frame of the damped Ly$\alpha$ system at $z_{\rm abs}$ = 0.18562, which is essential to studying the molecular and metal content of this system. The data have a resolution of  17--20 \kms, and S/N per resolution element of 15 -- 20. Processing of the COS data follows the steps outlined in \citet{Meiring2011} and \citet{Tumlinson2011}.

\subsection{Atomic and Molecular Hydrogen Gas  Modeling}
\label{h2col}

\begin{figure}
\begin{center}
\includegraphics[angle=90,scale=0.65]{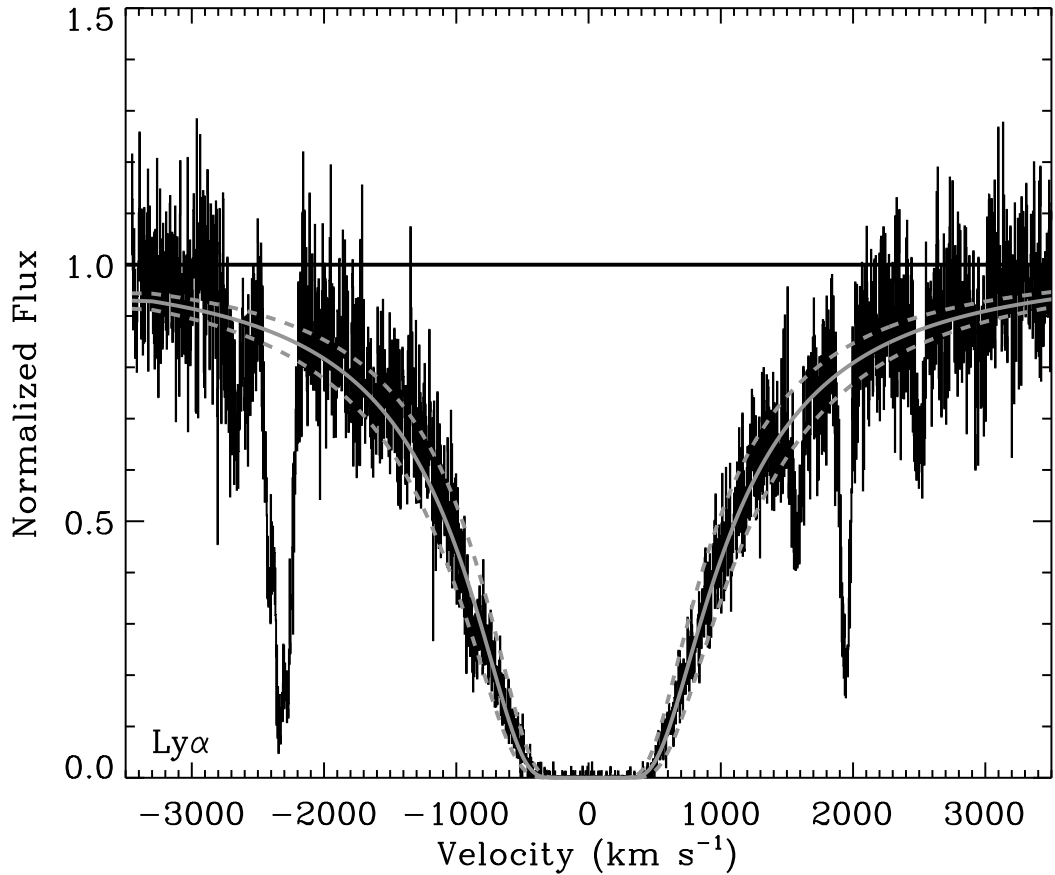}
\caption{Fit to the H\,I damped Ly$\alpha$ transition at $z_{\rm abs} = 0.18562$ towards QSO B0120$-$28, yielding log $N$(H\,I) = 20.50 $\pm$ 0.10. The solid line is the best fit profile; the dashed curves are the 1$\sigma$ uncertainties. \label{HI_fit}}
\end{center}
\end{figure}

From Voigt-profile fitting of a single absorption component to the $z_{\rm abs}$ = 0.18562 damped Ly$\alpha$ line we derive a total neutral hydrogen column density of log $N$(H\,I) = 20.50 $\pm$ 0.10. The fit with 1 $\sigma$ uncertainties is shown in Figure \ref{HI_fit}. The Ly$\beta$ transition falls on top of Ly$\alpha$ geocoronal emission, so we did not attempt to fit it.

\begin{figure}
\begin{center}
\includegraphics[angle=90,scale=0.65]{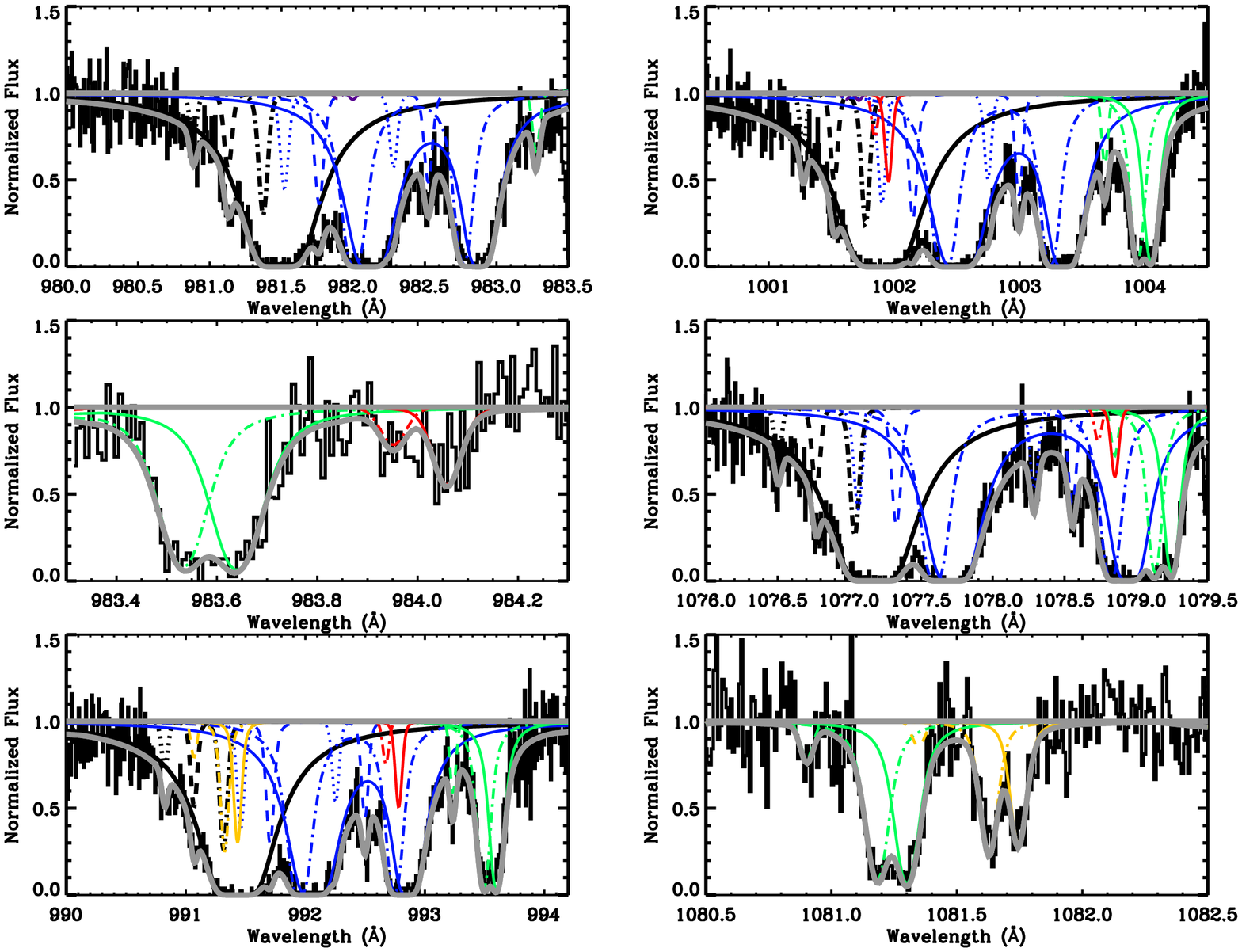}
\caption{Fits to a sample of  six \hmol~ spectral windows used to derive the column densities of the different $J$~levels are shown (in total, nineteen spectral windows were used in the fit). The overall best fit is overplotted in gray. Different colors are used for the different rotational levels as follows: black for $J =0$, blue for $J = 1$, green for $J = 2$, yellow for $J = 3$, and red for HD. Dotted lines are used for the component at $<v_1> = -170$ \kms, dashed for $<v_2> = -96$ \kms, dash-dot for $<v_3> = -20$ \kms, and solid lines for $<v_4> = +13$ \kms. Components 3 and 4 contain the bulk of the molecular gas in the $z_{\rm abs} = 0.18562$ DLA.  \label{H2_fit}}
\end{center}
\end{figure}


Absorption by \hmol~(up to $J$ = 3) is clearly detected in the $z_{\rm abs} = 0.18562$ DLA toward QSO\,B0120$-$28. Some lines from the $J$ = 4 and $J$ = 5 rotational levels may be be present but  are either blended with other lines or are too weak to be detected at the S/N of the data. Since most of the \hmol~gas is in the lower rotational levels, the contribution of the $J$ = 4 and 5 levels to the total $N$(\hmol) is negligible and will not be considered further.

Four absorption components are resolved in \hmol~at $<v_1> = -$170, $<v_2> = -$96, $<v_3> = -$20, and $<v_4> = +$13 km s$^{-1}$ (velocities given in relation to $z_{\rm abs} = 0.18562)$. HD is only detected in the latter two components (components 3 and 4, see below) where the bulk of \hmol~is concentrated. Because the \hmol~lines are blended with themselves and with other species, we could not determine $N$(\hmol) from a curve-of-growth. Instead, \hmol~column densities were derived by profile fitting a four component absorption model to the COS data. Fits to a sample of  six \hmol~ spectral windows used to derive the column densities of the different $J$~levels are shown in Figure \ref{H2_fit} (a total of 19 spectral windows were used to derive $N$(\hmol)). Different colors are used to identify the different rotational levels, while different line styles are used to identify the different absorption components. Column densities derived for each of the $J$ levels in each of the components are given in Table \ref{nsummary}. The total \hmol~column density for this system is log $N$(\hmol) = 20.00 $\pm$ 0.10, leading to log$N$(H) = 20.71 $\pm$ 0.07 and a mean molecular fraction $f$(\hmol) = 2$N$(\hmol)/(2$N$(\hmol) + $N$(H\,I)) = 0.39 $\pm$ 0.10. This is the highest molecular fraction of any DLA reported in the literature, and is similar to molecular fractions found in the disk of the Milky Way along diffuse and/or translucent cloud sightlines \citep[see][]{Rachford2002,Crighton2012}.

From the different $J$ levels in each \hmol~component we can derive excitation temperatures, using $T_{ij} = (\Delta E_{ij})/\kappa/ln[(g_j/g_i)(N(i)/N(j))]$, where $g_i$ is the statistical weight of the level $i$, $\kappa$ is the Boltzmann constant and $\Delta E_{ij}$ is the energy difference between levels $i$ and $j$.  In component 1 at $<v_1> = -$170 \kms, $T_{01}$ is consistent with 112 K and in component 2 at  $<v_2> = -$96 \kms, $T_{01}$ is consistent with 104 K. For component 3, at $<v_3> = -96$ \kms, the population of the different $J$ levels can be described by a single temperature, $T_{03} = 220 \pm 11$ K, indicating that in this component the \hmol~excitation is likely dominated by collisions. For component 4, at $<v_4> = +13$ \kms, we find $T_{01} = 65 \pm 2$ K and $T_{\rm 23} = 154 \pm 7$ K.

\subsection{Deuterated Molecular Hydrogen Column Density}
\label{N(HD)}

\begin{figure}
\begin{center}
\includegraphics[angle=90,scale=0.65]{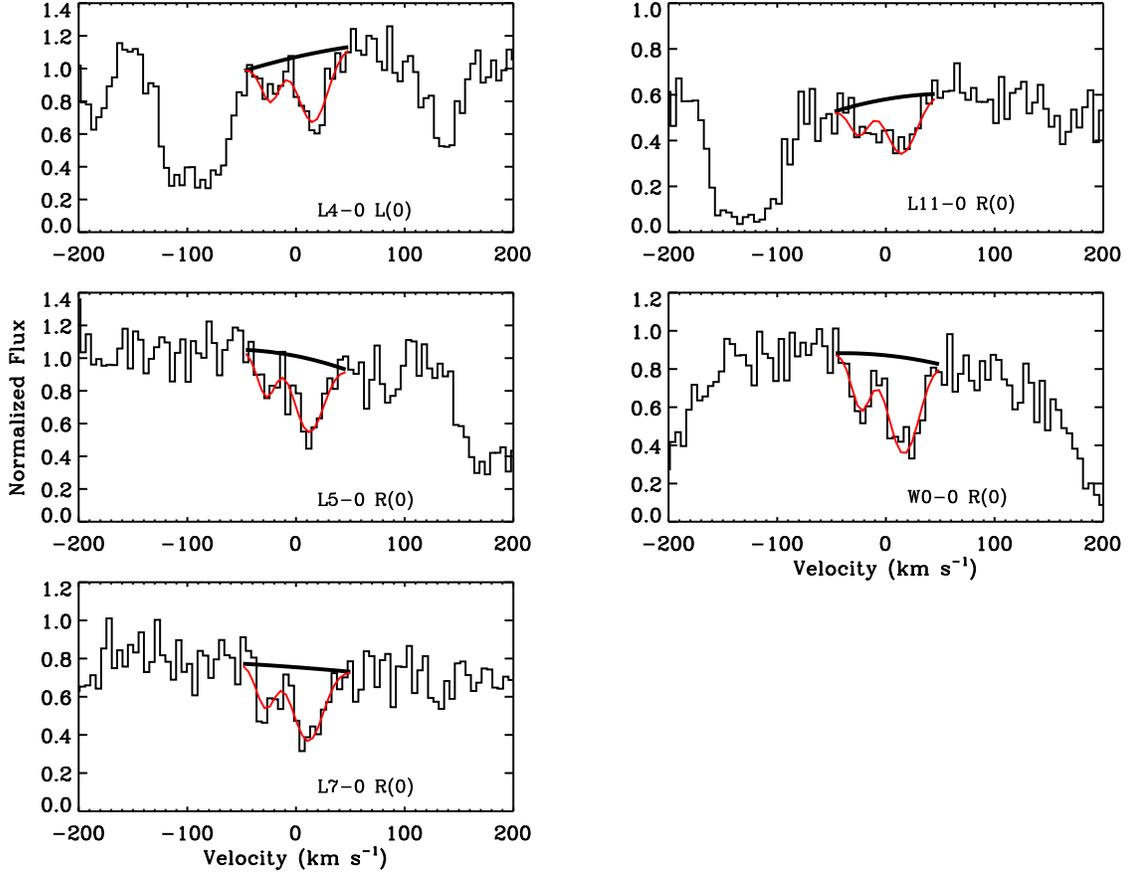}
\caption{Blend-free HD lines in the $z_{\rm abs} = 0.18562$ system towards QSO B0120-28. These  lines were used to derive the HD column density with profile fitting and apparent optical depth techniques (see discussion in \S\ref{N(HD)}). This figure shows a model with two components at  $<v_3> = -20$ \kms~ and $<v_4> = +13$ \kms~ fit to the data, yielding $N$(HD) = 14.68 $\pm$ 0.08.   \label{HDfit}}
\end{center}
\end{figure}

\begin{figure}
\begin{center}
\includegraphics[angle=90,scale=0.60]{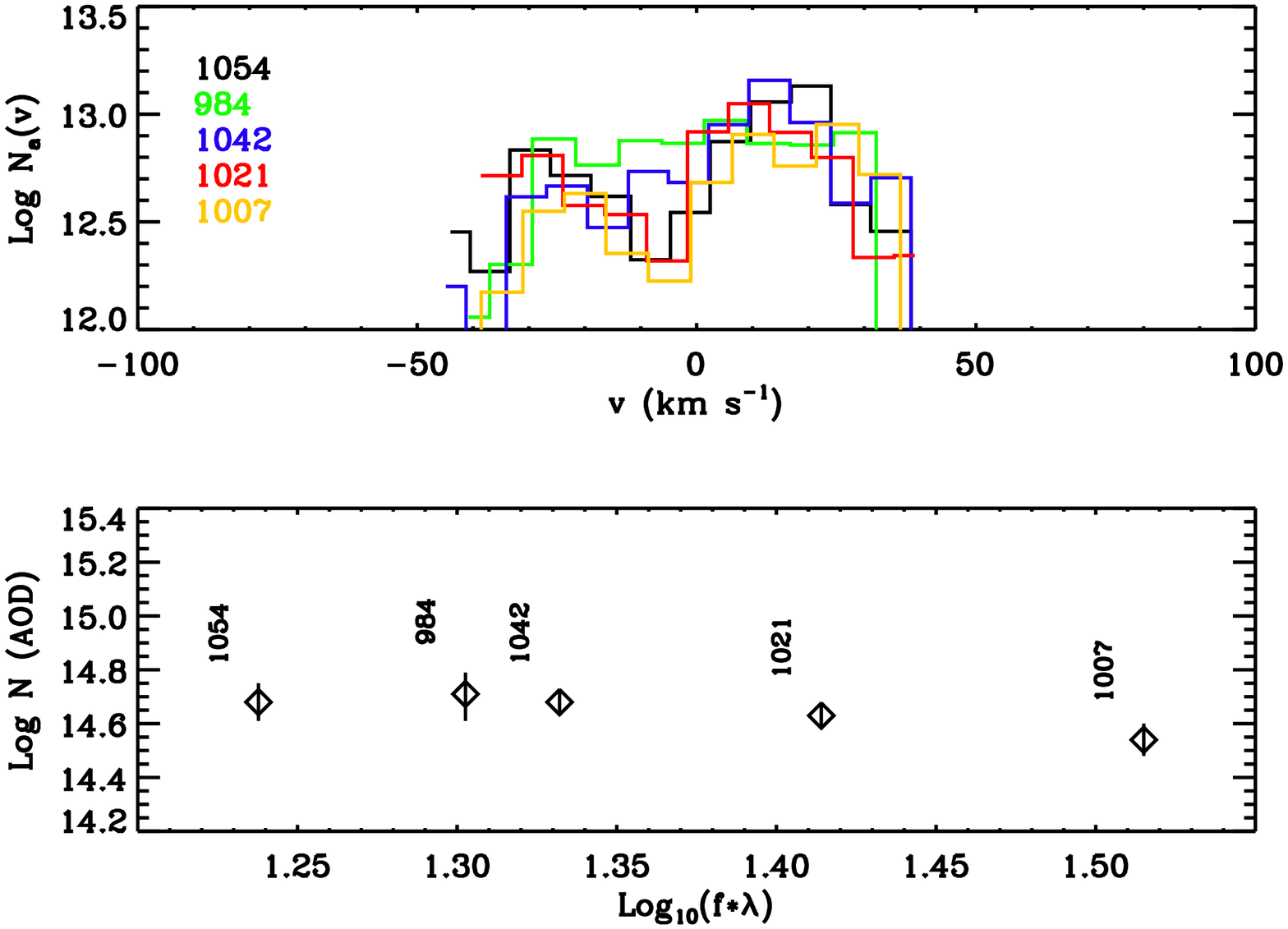}
\caption{{\it Top Panel}: Apparent column density, log $N_a(v)$, as a function of velocity for the blend-free HD lines. {\it Bottom Panel}: Column densities, derived from integrating the apparent column density profiles between $-48 < v < +45$ \kms, as a function of transition strength, log($f\times\lambda$). After correcting for unresolved saturation (see discussion in \S\ref{N(HD)}) the adopted total column density is log $N$(HD) $= 14.82 \pm 0.15$.   \label{AOD}}
\end{center}
\end{figure}

For HD ($J = 0$), the Lyman L3-0 through L8-0, L11-0, and L12-0 bands, as well as the W0-0 Werner band, are detected in two components at $<v_3> = -20$ and $<v_4> = +13$ \kms~(see Figure \ref{HDfit}).  Profile fitting of these transitions with a two component model yields log $N$(HD) = 14.14 $\pm$ 0.10 and $b$ = 4 $\pm$ 2 \kms~for the component at $-20$ \kms and log $N$(HD) = 14.53 $\pm$ 0.10 with $b$ = 13 $\pm$ 3 \kms~for the component at $+13$ \kms.  Figure \ref{HDfit} shows the results of the fit to the HD lines that are free from blends. The total column density derived with profile fitting is log $N$(HD) = 14.68 $\pm$ 0.08. The Doppler parameters derived with profile fitting are higher that the 1 -- 2 \kms~found in higher $z$ HD bearing DLAs \citep[see e.g.][]{Tumlinson2010} and likely indicate there is unresolved structure in the lines. To assess the impact of unresolved structure in $N$(HD) we use HD lines that are free from blends (shown in Figure \ref{HDfit}) and  plot in Figure \ref{AOD} the log of the apparent optical depth (AOD) as a function of velocity, log $N_{a}(v)$, for the blend-free HD transitions \citep[for more information on the AOD technique see][]{Savage1991}. Because the HD  lines are not fully resolved we determine the apparent column density for each transition by integrating the apparent column density profile over the two HD components, between $-48 < v < +45$ \kms. The bottom panel of Figure \ref{AOD} shows the derived apparent column density as a function of transition strength given by log ($f \lambda$) (see Table \ref{hdmeas} for a summary of the column densities). In the case of unresolved saturated structure the apparent column density profiles, $N_a$, of two lines that differ in strength by at least a factor of two should show significant departures from one another. In our case the profile of the stronger line ($\lambda$1007, yellow, in the top panel of Figure \ref{AOD}) falls below that of the weaker line ($\lambda$1054, black) indicating the presence of unresolved saturation (the two HD lines differ in strength by a factor of 1.9). We use  formula 13 of \citet{Savage1991} to correct log $N$(HD)$_{\rm 1054}$ by a factor of 0.14 dex, the difference between log $N$(HD)$_{\rm 1054}$ = 14.68 $\pm$ 0.07 and log $N$(HD)$_{\rm 1007}$ = 14.54 $\pm$ 0.06, to adopt log $N$(HD) = 14.82 $\pm$ 0.15. The uncertainty of 0.15 dex takes into account some of the uncertainties in deriving saturation corrections. The total column density derived with profile fitting, log $N$(HD) = 14.68 $\pm$ 0.08, is lower than our adopted value due to saturation, but provides a consistency check with the column density derived with the AOD technique.

\subsection{Metallicity, Depletion, and Dust Content}

In the $z_{\rm abs} = 0.18562$ DLA system we detect also absorption by C\,II, C\,III, N\,I, N\,II, O\,I, Si\,II, Si\,III, P\,II, S\,II, Ar\,I, and Fe\,II. No absorption by N\,V, O\,VI, or Si\,IV is detected. While the analysis of the metal content of the DLA will be discussed elsewhere (Oliveira et al., in prep) here we focus on the elements Si, S, and Fe that allow us to constrain the metallicity, depletion, and dust content of the DLA. 

\begin{figure}
\begin{center}
\includegraphics[angle=90,scale=0.6]{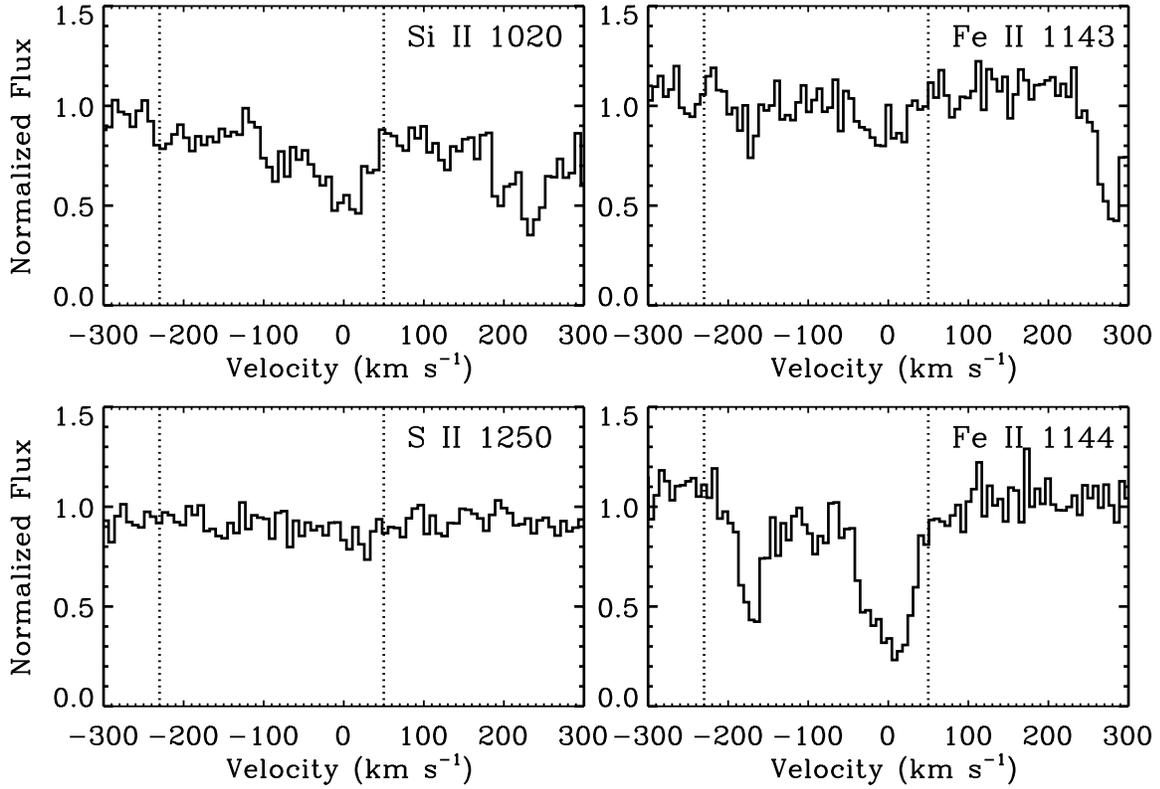}
\caption{Some of the metals present in the $z_{\rm abs} = 0.18562$ DLA are shown. Dotted vertical lines indicate the velocity limits (in the DLA frame of rest) over which the absorption occurs. The profiles were integrated between those velocity limits when determining the column densities with the AOD technique. The data has been binned by 3, roughly half the size of a resolution element, for display purposes only.\label{metals}}
\end{center}
\end{figure}

Figure \ref{metals} shows the absorption profiles of Si II (1020.6989 \AA), S II (1250.5840 \AA), and Fe II (1143.2260 \AA, 1144.937 \AA). Dotted vertical lines indicate the velocity limits (in the DLA rest frame) of the absorption associated with the $z_{\rm abs} = 0.18562$ DLA. We use the AOD technique to determine the column densities of Si\,II, S\,II, and Fe\,II, and we integrate the apparent column density profiles over the velocity ranges marked by vertical dotted lines in Figure \ref{metals}. We find log $N$(S II) = 14.67$^{+0.13}_{-0.20}$ (the $\lambda\lambda$1253, 1259 transitions of S\,II are blended with absorption not related to the DLA and so cannot be used to constrain $N$(S\,II)), log $N$(Si\,II) = 15.07 $\pm$ 0.05, and log $N$(Fe\,II) = 14.68$^{+0.08}_{-0.06}$ (from the weaker $\lambda$1143 transition). Saturation effects are negligible for the S\,II and Fe\,II transitions used to derive the column densities; S\,II $\lambda$1250 is barely detected and $N$(Fe\,II) derived from the $\lambda$1143 transition agrees within 1 $\sigma$ with that derived from $\lambda$1144 which is a factor of $\sim$6 stronger in oscillator strength.  We make the assumption that $N$(S) $\sim$ $N$(S\,II) and $N$(Fe) $\sim$ $N$(Fe\,II) because no other ionization stages of S or Fe are detected.  We use the solar abundances from \citet{Asplund2009} together with log $N$(H) = 20.71 $\pm$ 0.07 from \S\ref{h2col} and [X/H] = log [$N$(X)/$N$(H)] - log [$N$(X)/$N$(H)]$_{\odot}$ to derive [S/H] =  -1.19$^{+0.15}_{-0.21}$ and [Fe/H] = -1.48$^{+0.11}_{-0.09}$. 
In the case of Si we need to take into account Si\,III and so  we use the AOD technique to place a lower limit on Si\,III, log $N$(Si\,III) $> 13.71$, using the saturated $\lambda$1206.5 transition. Because $N$(Si\,III) is not well constrained, [Si/H] $\geq$ -1.15 $\pm$ 0.09.  The metallicities derived from S and Si, 0.06--0.07$Z_{\odot}$,  are similar even if Si\,III is not fully accounted for. Fe is prone to depletion and so the metallicity implied is only 0.03$Z_{\odot}$.
The metallicities derived from S and Si are similar to the metallicities of  $z < 0.4$ DLA and sub-DLA systems  studied by \citet{Battisti2012} who found values  in the range 0.08--2$Z_{\odot}$. The study by \citet{Battisti2012} is one of the few that exists of relative abundances in DLAs at low $z$.	 

It is often assumed that DLA systems are not affected by ionization corrections due to the high H\,I column densities probed. This would be true if all the H\,I was contained in a single cloud - that is not the case for the DLA studied here where four absorption components are resolved in \hmol. The second  ionization potential of sulphur, 23.4 eV, is higher than that of H$^0$, and so S II can exist in regions where hydrogen has been ionized, implying that the mean metallicity derived above is an upper limit to the true metallicity if a large fraction of ionized gas was present in the DLA. The metallicity derived here is only a mean value; it is possible that the metallicity in the components containing the bulk of the molecular gas is higher than the mean value. An upper limit to the metallicity of the strongest \hmol~components, i.e. components 3 and 4, can be derived assuming $N$(H) = 2$\times N$(\hmol) and using log $N$(S) = 14.59 $\pm$ 0.10 between -60 $\le v \le +50$ \kms. The upper limit derived, [S/H] $< -0.86$, is  a factor of $\sim$two higher than the overall metallicity of the DLA, [S/H] =  -1.19$^{+0.15}_{-0.21}$.  A similar upper limit, [S/H] $<$ -0.81 is derived for component 4 alone, using N(S) = 14.59 $\pm$ 0.10 between -60 $\le v \le +50$ \kms. These results indicate that the metallicity in the components containing the bulk of \hmol~could be higher than the overall metallicity of the DLA by at most a factor of $\sim$2.

 

 The depletion factor of Fe can be determined from [X/Fe] = log [$N$(X)/$N$(Fe)] - log [$N$(X)/$N$(Fe)]$_{\odot}$, using S as the reference element that is little or unaffected by depletion, leading to a mild depletion of Fe, [S/Fe] = +0.29$^{+0.15}_{-0.21}$.  This is again a mean depletion factor for the gas in the DLA. A more realistic depletion factor can be determined for the gas containing the bulk of the molecular content by determining [S/Fe] for the velocity interval corresponding to the absorption seen in components 3 and 4 in Table \ref{nsummary}. Integrating the apparent column densities of S\,II ($\lambda$1250) and Fe\,II ($\lambda$1143) over the interval $-60 < v < +50$ \kms~leads to log $N$(S\,II) = 14.59 $^{+0.10}_{-0.13}$ and log $N$(Fe\,II) = 14.47 $^{+0.06}_{-0.08}$. The depletion factor of Fe, [S/Fe] = 0.42 $^{+0.12}_{-0.15}$, is higher in the components containing the bulk of the molecular gas than the average value for the DLA.
  
 A good indicator of the dust content in the DLA is the dust-to-gas ratio $\kappa_{\rm X} = 10^{\rm [X/H]}(1 - 10^{\rm [Fe/X]})$ \citep{Ledoux2003}. Using again S as the reference element unaffected by dust depletion effects leads to a low average dust content of $\kappa_{\rm X}$ = 0.03 or log ($\kappa_{\rm X}$) = -1.5.  Similar to the Fe depletion factor, we suspect that the dust-to-gas ratio in the components containing the bulk of the molecular gas is higher than the average value derived above.

\section{Discussion}
\label{discussion}

Several quantities derived here indicate that the properties of the DLA system at $z = 0.18562$ in the QSO B0120-28 sightline are similar to those in cool clouds in the disk of the Galaxy. The excitation temperature derived for the component containing the bulk of the gas, $T_{01} = 65 \pm 2$ K,  is similar to the values found by \citet{Savage1977} for the Galactic disk with {{\it Copernicus},  $T_{01} = 77 \pm 17$ K, while from a sample of several DLAs at $z > 1.9$ \citet{sri2005} derived a mean temperature  $T_{01} = 153 \pm 78$ K. The mean molecular fraction, $f_{\rm H_2}$ = 0.39 $\pm$ 0.10, is the highest value reported in the literature for DLAs and is consistent with the disk sightlines in the HD survey by \citet{Lacour2005}, which have typical molecular fractions between 0.2 and 0.6. The overall metallicity of the DLA, [S/H] =  -1.19$^{+0.15}_{-0.21}$, is also similar to that of cool clouds in the disk of the Milky Way; \citet{Savage1992} found [Si/H] $= -1.31$ for $\zeta$ Oph, which also has a large molecular fraction, $f_{\rm H_2}$ $\sim 0.6$.

\begin{figure}
\begin{center}
\includegraphics[angle=90,scale=0.70]{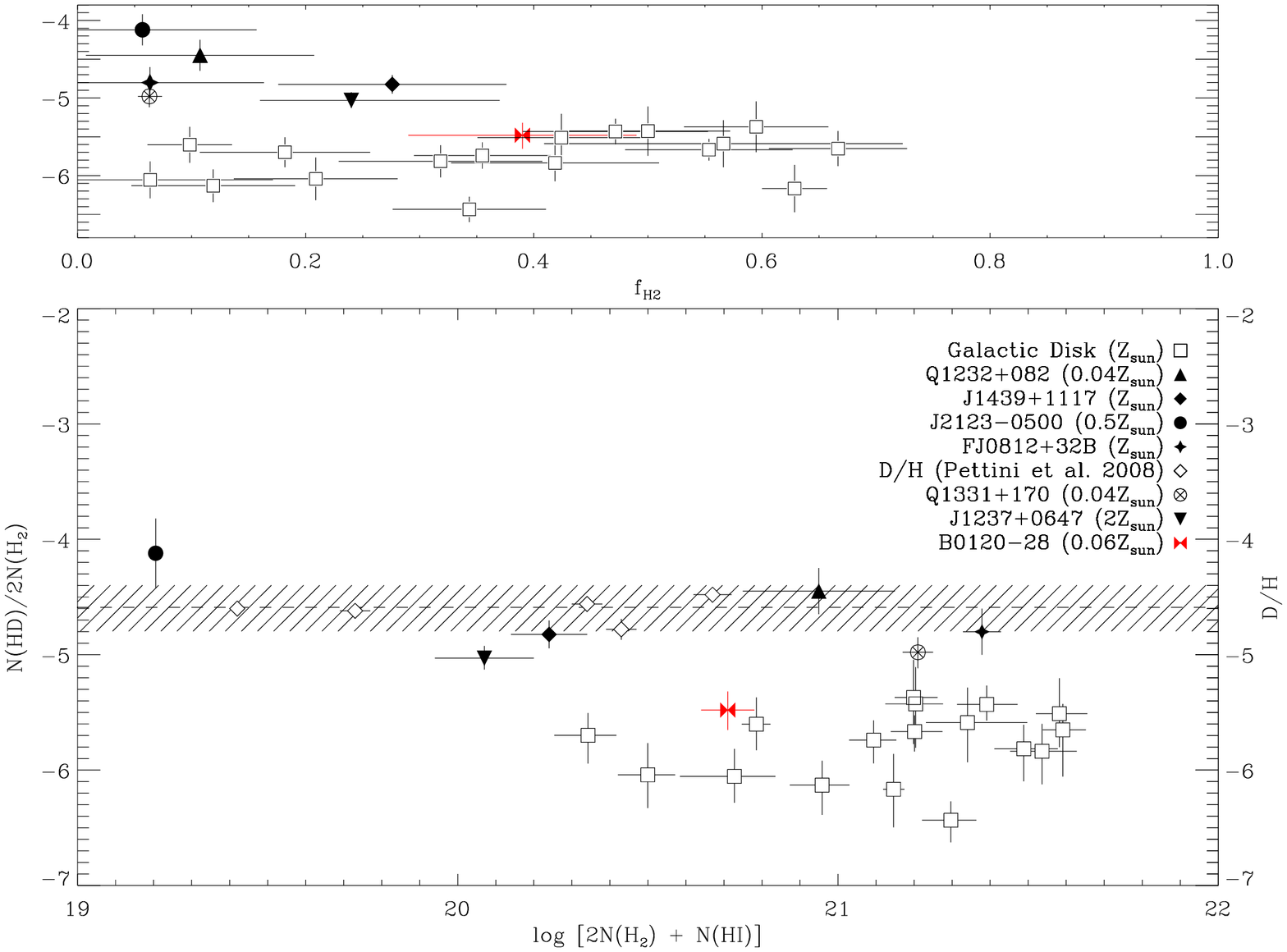}
\caption{Comparison of our new HD measurement (red bowtie symbol) with measurements in the literature as a function of the fraction of molecular hydrogen ({\it top panel}) and the total hydrogen column density ({\it bottom panel}): The DLA systems from \citet[][filled triangle]{Ivanchik2010}, \citet[][filled diamond]{Noterdaeme2008HD}, \citet[][circle with cross]{Balashev2010}, and \citet[][filled star]{Tumlinson2010}, and the sub-DLA systems from \citet[][inverted triangle]{Noterdaeme2010} and \citet[][filled circle]{Tumlinson2010}. Measurements in the Galactic disk from \citet{Lacour2005} are marked with open squares. The cross-hatched area refers to the axis at right and marks the CMB-derived D/H ratio from \citet{Dunkley2009}. Open diamonds are D/H measurements in high-redshift DLAs from the compilation by \citet{Pettini2008}. \label{JTfig}}
\end{center}
\end{figure}

As with Galactic disk sight lines of similar HD and \hmol~column densities, one might expect to detect other species in the $z_{\rm abs} =$ 0.18562 system, such as Cl\,I, CO, and C\,I.
Cl\,I is a tracer of cold \hmol~gas \citep{Sonnentrucker2006} and one would expect to detect it at the high \hmol~column densities derived for the $z_{\rm abs} = 0.18562$ DLA host; however no Cl\,I is  detected due to metallicity effects.  From the $\lambda$1347 transition we place an upper limit of log $N$(Cl\,I) $<$ 13.00 with the AOD technique. Using the Cl\,I/\hmol~ relationship derived by \citet{Sonnentrucker2002} and \citet{Moomey2012} for the Milky Way, Cl\,I/\hmol~$\sim 4\times10^{-7}$, we predict log $N$(Cl\,I) = 13.60. However, when we take into account the low metallicity of the DLA system, [S/H] = -1.2 (corresponding to $Z = 0.07Z_\odot$), together with the Cl abundance in nearby H\,II regions \citep{Garcia2007,Asplund2009} as a proxy for the solar abundance, log $\epsilon_{\rm H} = 5.32 \pm 0.07$, we predict a column density log $N$(Cl\,I) $\sim$ 12.14, which is well below our detection limit.

CO is another tracer of cold gas that has been detected in some DLA systems \citep[see e.g.][]{Sri2008,Noterdaeme2009}. Using the CX (0-0) band oscillator strength from \citet{Federman2001} we derive an upper limit to the column density of CO, log $N$(CO) $< 13.03$.  This value is similar to the value of log $N$(CO) $\sim 13$ found using the CO/\hmol$\sim 10^{-7}$ conversion factor,  for diffuse clouds in the Galaxy \citep{Burgh2007}.  Metallicity effects in the DLA host would work to lower the CO/\hmol~conversion ratio, leading to $N$(CO) below our upper limit.

  Figure \ref{JTfig} compares our new HD measurement (red  symbol) with measurements in the literature, as a function of the fraction of molecular hydrogen (top panel) and the total hydrogen column density (bottom panel), and with Galactic sight lines from \citet{Lacour2005}.   The HD to \hmol~ratio, log(HD/2\hmol) = -5.5 $\pm$ 0.2, is similar to disk sight lines and is lower that what has been found in higher redshift DLAs known to contain HD. All high-redshift DLAs have $ -5.03 \pm 0.10 <$ log(HD/2\hmol) $< -4.10 \pm~^{0.22}_{0.24}$, well above the Galactic disk value and consistent with the cosmological D/H ratio \citep[cross-hatched area in Figure \ref{JTfig} from][]{Dunkley2009} and with the D/H ratios seen in high-redshift DLAs \citep[open diamonds in Figure \ref{JTfig} from the compilation by][]{Pettini2008}.    
  
The low D/H ratio derived from HD/2\hmol~is likely not a consequence of D-astration in stars. The wide scatter in the metallicity of the HD systems shown in Figure  \ref{JTfig} and its lack of relationship with HD/2\hmol~both argue against astration: some of the systems with the highest HD/2\hmol~also have the highest metallicity. From D/H measurements in high redshift QSOs \citet{Pettini2012} derive a mean primordial value (D/H)$_{\rm P}$ = (2.60 $\pm$ 0.12)$\times10^{-5}$. In the local disk, within 1 kpc of the Sun, direct measurements vary between D/H = (5.0 $\pm$3.4)$\times10^{-5}$ and D/H = (2.2 $\pm$ 0.9)$\times10^{-5}$ \citep{Linsky2006}. It  has been argued that  the wide range of D/H measurements could be due to turbulent mixing \citep{Bell2011}, which could either decrease or increase D/H compared to the "true" elemental ratio or to depletion \citep{Linsky2006}, which would lower the D/H ratio compared to the "true" value. While probably both mechanisms, turbulent mixing and depletion, contribute to the observed D/H in the local disk, the picture made by these measurements is consistent with cosmic evolution models, that predict that deuterium astration from its primordial value is at most a factor of $\sim 2-3$ \citep{Romano2006,Olive2012}. 

Similarly to Galactic sightlines, the D/H ratio derived from HD/2\hmol~in the $z = 0.185262$ DLA, D/H $= (3.2 ^{+1.9}_{-1.2})\times10^{-6}$, is not expected to reflect the true elemental ratio, unless all the deuterium is in molecular form.  In the Galactic disk HD exists only in the dense parts of clouds given that self-shielding of HD becomes significant only at higher extinction than for \hmol, due to the lower abundance of deuterium compared to hydrogen. As pointed out by \citet{Tumlinson2010},  clouds with lower molecular fractions should have more of their HD dissociated and should lie below the Galactic disk points (see bottom panel of Figure \ref{JTfig}).  

Given the considerations above the question becomes then why is D/H derived from HD/2\hmol~in high redshift DLAs with low molecular fraction so close to the primordial value? Is there a point in the evolution of these systems across cosmic time when the conditions change in such a way that these systems start resembling the cold clouds in the Milky Way disk? As \citet{Tumlinson2010} point out, the uncertainty in chemical reactions, dust properties, and local conditions in the high redshift systems  prevent us from understanding the conditions that lead to the enhancement of HD/\hmol. Only the discovery of more similar systems, coupled with additional theoretical work, will allow us to understand the interstellar chemistry in high redshift DLAs.

We have presented the detection of HD in a DLA system at $z_{\rm abs} = 0.18562$ toward QSO\,B0120-28.
As discussed above, the characteristics of the cold gas in this system are unlike those found in other high DLA systems but are consistent with what is observed in dense clouds in the Milky Way. Using ground based imaging and spectroscopic data we have identified a galaxy at $z = 0.18562$~that lies at an impact parameter of $\sim$70 Kpc ($\sim$ 23 \arcsec) from the QSO sightline (assuming $\Omega_{\lambda} = 0.70$, $\Omega_{\rm m} = 0.30$, and $H_{\rm 0} = 0.70$ \kms). The gas probed by \hmol~and HD could be associated with the extended  disk of this galaxy or could be remnant tidal debris from a previous interaction between the DLA host and another galaxy. A detailed discussion of the DLA host and its metal content will be presented in a follow-up paper (Oliveira et al., in prep).

\acknowledgments


\clearpage

\begin{deluxetable}{lcccc}
\tabletypesize{\small}
\setlength{\tabcolsep}{0.02in} 
\tablecaption{\hmol~Column Densities (log)  \label{nsummary}}
\tablewidth{0pt}
\tablehead{
\colhead{Component} & \colhead{ (1)} & \colhead{(2) } & \colhead{(3)} & \colhead{(4)}\\
\colhead{ $<v>$} & \colhead{ -170 \kms} & \colhead{ -96 \kms} & \colhead{ - 20 \kms} & \colhead{ +13 \kms}}
\startdata
$N$(\hmol)~($J  = 0$) &  16.14 $\pm$ 0.14 & 16.80 $\pm$ 0.13 & 16.81$^{+0.87}_{-0.22}$ & 19.72 $\pm$ 0.02 \\
$N$(\hmol)~($J  = 1$) &  17.23 $\pm$ 0.08 & 17.45 $\pm$ 0.08 & 18.91 $\pm$ 0.07             & 19.53 $\pm$ 0.03 \\		
$N$(\hmol)~($J  = 2$) & \ldots                        &        14.55$^{+0.16}_{-0.08} $                      & 18.32 $\pm$ 0.05             & 18.40 $\pm$ 0.04	\\
$N$(\hmol)~($J  = 3$) & \ldots                        &		 14.21 $\pm$ 0.10			    & 17.73 $\pm$ 0.05            & 17.60 $\pm$ 0.05 \\	
\hline
$N$(\hmol)  (total)       & 17.26 $\pm$ 0.08  &  17.54 $\pm$ 0.07  & 19.03 $\pm$ 0.06   & 19.95 $\pm$ 0.02 \\
\enddata
\end{deluxetable}

\begin{deluxetable}{lcccc}
\tablecaption{Summary of HD Measurements \label{hdmeas}}
\tablewidth{0pt}
\tablehead{
\colhead{Line} & \colhead{ $\lambda_0$ (\AA)\tablenotemark{a}} & \colhead{$f$\tablenotemark{b}} & \colhead{W$_{\lambda}$ (m\AA)} & \colhead{Log $N_{a}$ (cm$^{-2}$)}}
\startdata
L4-0R(0)	& 1054.29 & 0.0164 & 65 $\pm$ 10 &   14.68 $\pm$ 0.07	\\
L5-0R(0)	& 1042.85 & 0.0206 & 77 $\pm$ 7 &   14.68 $\pm$ 0.05 \\
L7-0R(0)	& 1021.46 & 0.0254 & 79 $\pm$ 8 &   14.63 $\pm$ 0.05 \\
L11-0R(0)	& 984.013 & 0.0204 & 75 $\pm$ 14 & 14.71 $\pm$ 0.09 \\
W0-R(0)	& 1007.29 & 0.0325 & 79 $\pm$ 10 & 14.54 $\pm$ 0.06 \\
\enddata
\tablenotetext{a}{From \citet{Ivanov2008} except for L11-0R(0) which is from \citet{Abgrall2006}.}
\tablenotetext{b}{From \citet{Abgrall2006}.}
\end{deluxetable}

\clearpage

\end{document}